\documentclass{PoS}

\usepackage{graphbox}

\newcommand{\Nmu}{N_\mu}
\newcommand{\Xmax}{X_{\rm max}}
\newcommand{\Lhad}{\Lambda_{\rm had}}
\newcommand{\Ehad}{E_{\rm had}/E_0}

\def\sib{{\textsc{Sibyll 2.3c}}}
\def\qgs{\textsc{QGSJet-\,II.04}}
\def\epos{\textsc{epos-lhc}}

\title{Probing the high energy spectrum of neutral pions in ultra-high energy proton-Air interactions}

\ShortTitle{Tail of the $\Nmu$ distribution}

\author{Lorenzo Cazon, \speaker{Ruben Concei\c{c}\~{a}o}\thanks{Acknowledges the financial support of Funda\c{c}\~ao para a Ci\^encia e Tecnologia, FCT-Portugal (DL57/2016/cP1330/cT0002).} , Miguel Martins, Felix Riehn\\
        Laborat\'orio de Instrumenta\c{c}\~{a}o e F\'isica Experimental de Partículas - LIP and Instituto Superior T\'ecnico - IST, Universidade de Lisboa - UL, Portugal\\
        E-mail: \email{ruben@lip.pt}}

\abstract{The interaction of ultra-high energy cosmic rays with the atmosphere nuclei has long been seen as a unique opportunity to study hadronic interactions above energies attainable by accelerators. However, so far the multiparticle production properties of the first interaction have been difficult to assess as they are masked by the many interactions that outline the shower development.
In this work,  we demonstrate, that relevant properties of the ultra-high energy first interaction can be accessed through the analysis of the shower-to-shower distribution of muons arriving at the ground. In particular, it is shown that the slope of the low-tail of the number of muon distribution measured at the ground is a direct link to the high energy spectrum of neutral pions produced in the interaction of the primary protons.
In this presentation, it will also address the experimental feasibility of such measurements and their connection with physical quantities being currently measured at the Large Hadron Collider.}

\FullConference{36th International Cosmic Ray Conference -ICRC2019-\\
		July 24th - August 1st, 2019\\
		Madison, WI, U.S.A.}

\begin{document}

\section{Introduction}

Ultra-high energy cosmic rays (UHECRs) have been long seen as an opportunity to study the high-energy Universe and access hadronic interactions for center-of-mass energies above the 100 TeV scale.
However, the degeneracy between the primary composition and hadronic interactions together with the complex behavior of the Extensive Air Shower (EAS) development has undermined our ability to explore the properties of the first interaction, the exception being the measurement of the proton-Air cross-section.
Muons constitute one of the best links to the hadronic activity within the shower, and for that, there has been a large enterprise to understand their behavior both from the theoretical (see for instance \cite{MuonTransport, MPDpheno}) and from the experimental point of view, e.g.~\cite{AugerMuon, TAMuon}.
In this work, we investigate what drives the muon number distribution at the ground. We prove, through simulation, that while the average of the distribution depends on the sum of all hadronic interactions, its shape depends essentially on what happens in the very first interaction of the shower.
This manuscript is organized as follows: in section~\ref{sec:2} we discuss the origin of the fluctuations of the $\Nmu$ distribution, while in section~\ref{sec:3} we investigate the exponential tail that appears in the distribution for proton showers; In the sub-section~\ref{sec:3.1} we discuss how to use this tail to access the energy flow between the hadronic and electromagnetic (e.m.) component in the first interaction unequivocally and in sec.~\ref{sec:3.2} we discuss that the muon tail can also be used to access the high-energy tail of the neutral pion energy spectrum; The obtained results are summarized in section~\ref{sec:conclusions}.

\section{Shower-to-shower distribution of the EAS muon content}
\label{sec:2}

\begin{figure}[t]
\centering
\includegraphics[width=0.5\textwidth]{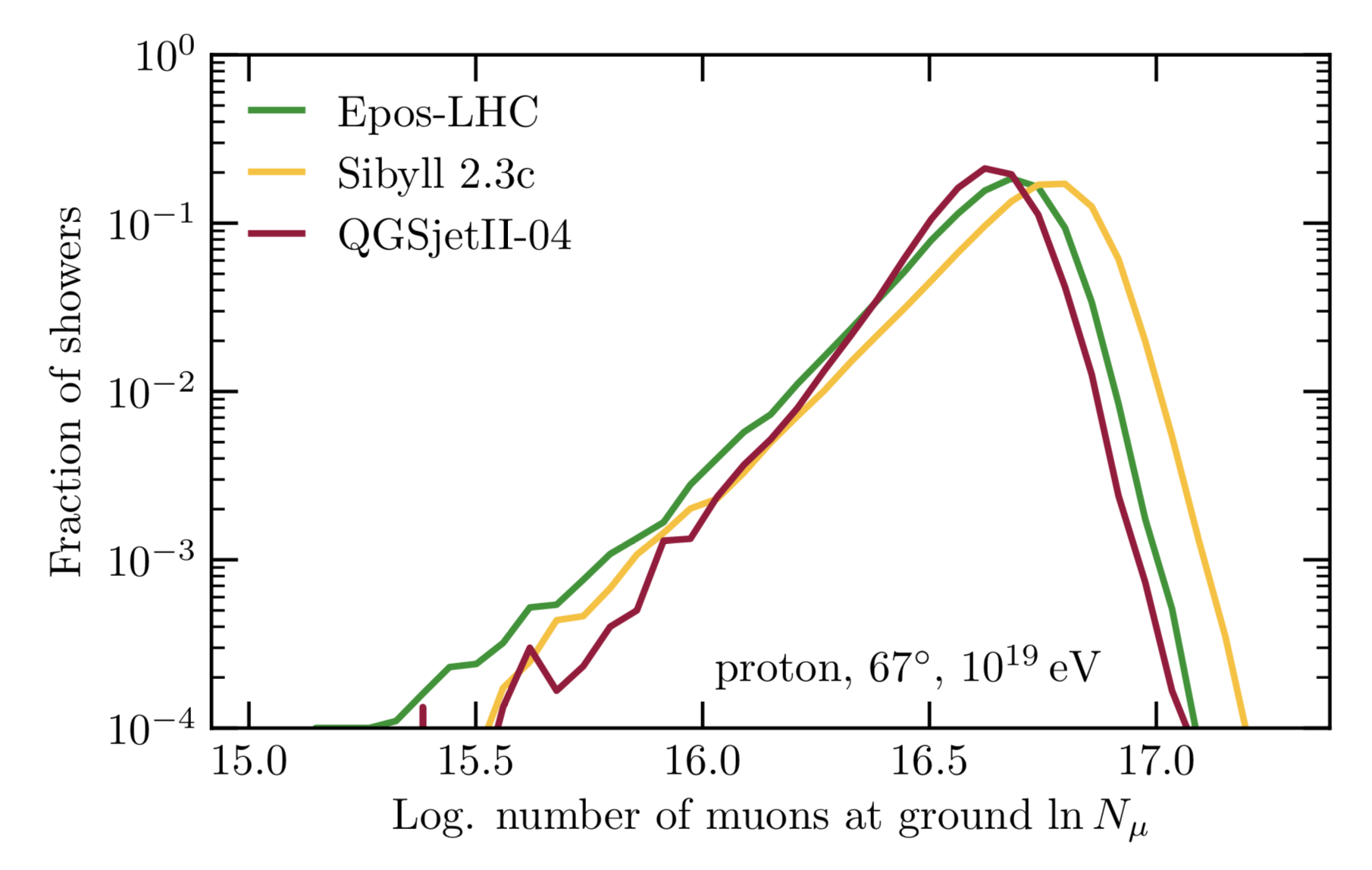}
\hfil
\centering
\includegraphics[width=0.29\textwidth]{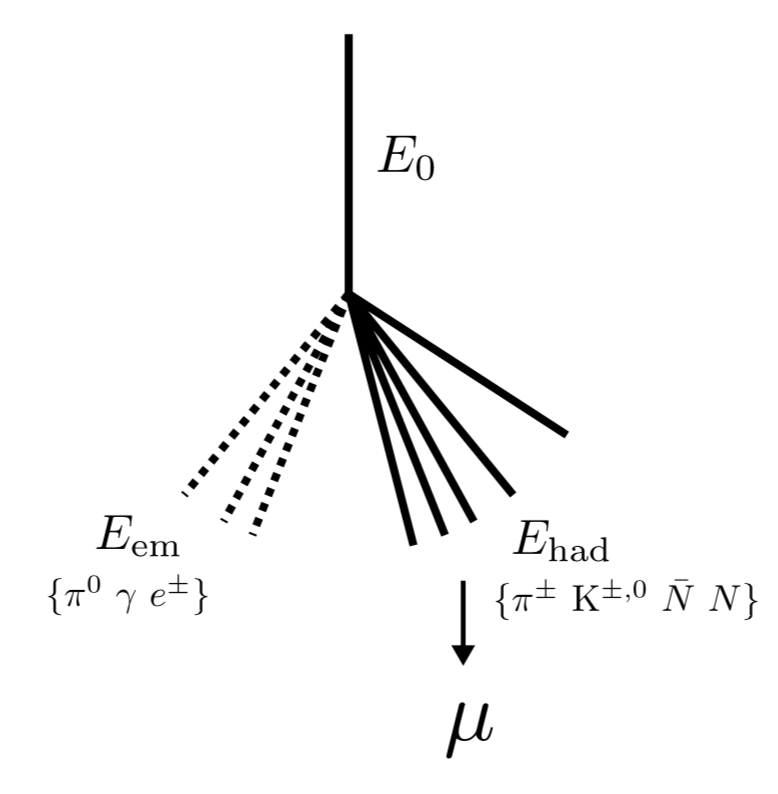}
\caption{(left) Distribution of the number of muons at the ground in simulated EAS induced by protons with $E = 10^{19}\,$eV and $\theta = 67^\circ$. The curve were generated with CONEX and so the muon energy threshold is $1\,$GeV. (right) Scheme of the energy flow in the first interaction into the electromagnetic and hadronic component of the shower.}
\label{fig:scheme}
\end{figure}

\begin{figure}[t]
\centering
\includegraphics[width=0.45\textwidth]{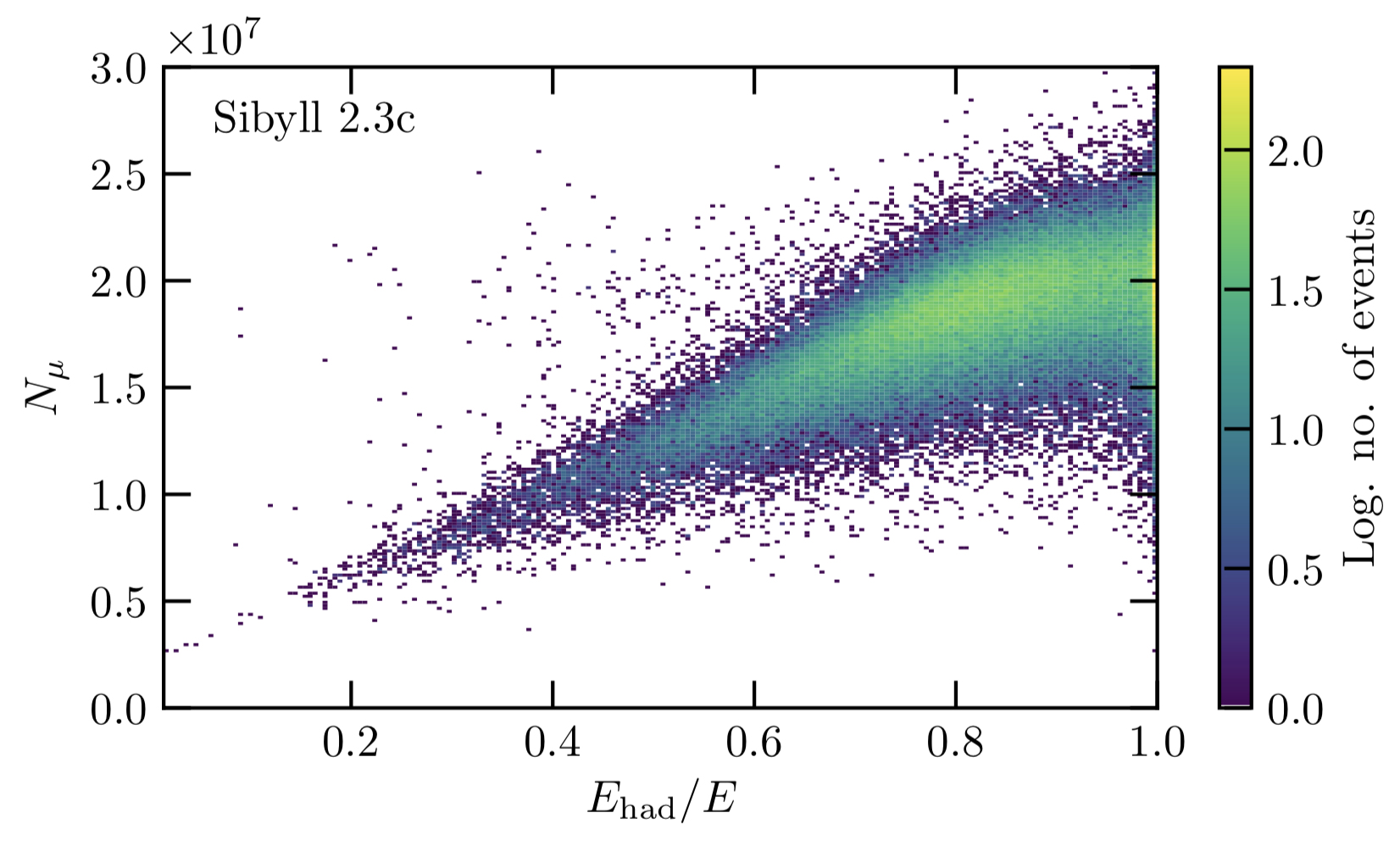}
\hfil
\includegraphics[width=0.45\textwidth]{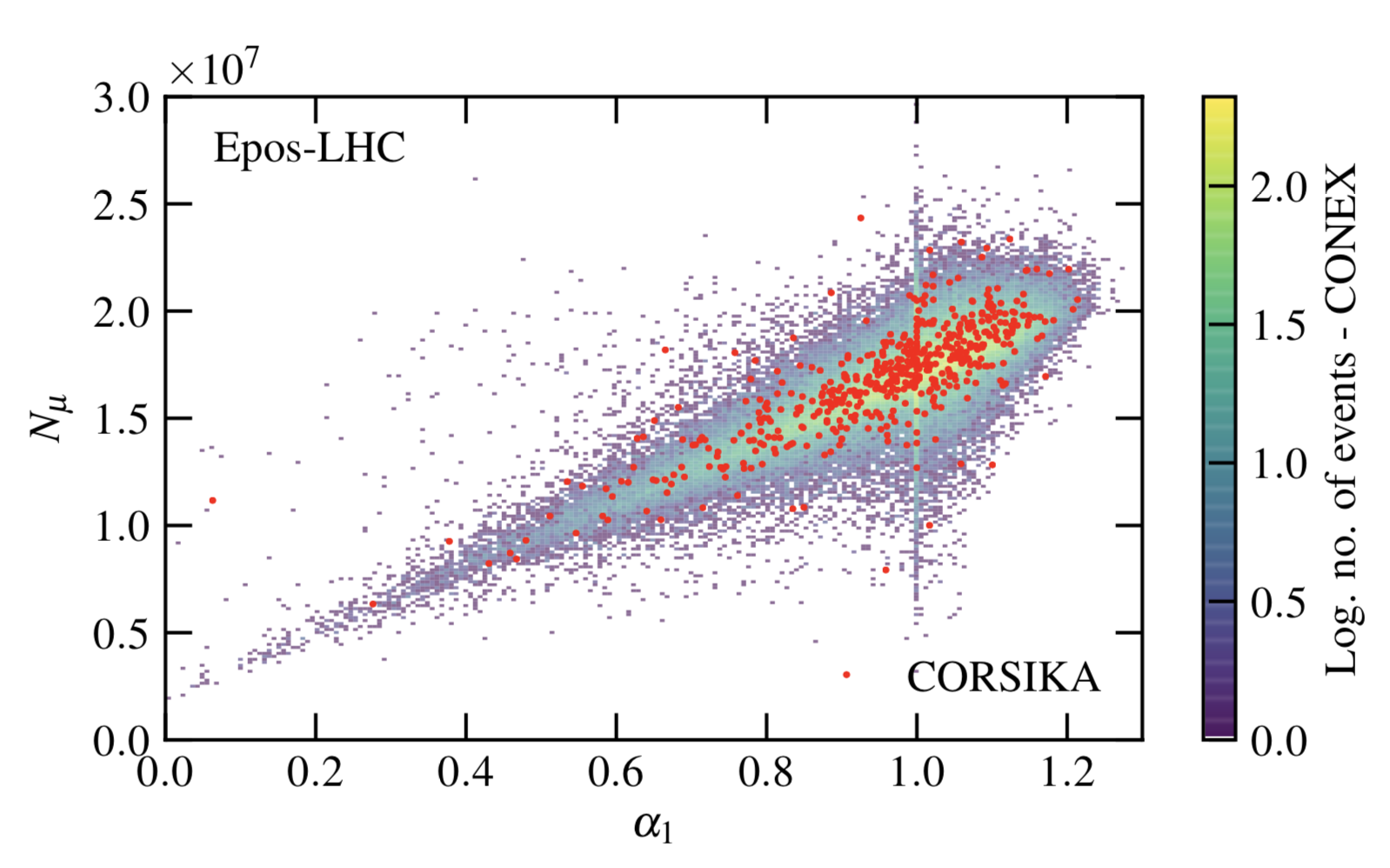}
\caption{(left) Distribution of the hadronic energy, $\Ehad$, and $\Nmu$ and (right) distribution of $\alpha_1$ versus $\Nmu$. Both plots were generated with showers induced by protons having an energy of $E = 10^{19}\,$eV and a zenith angle of $\theta = 67^\circ$.}
\label{fig:corr_alpha_nmu}
\end{figure}

It is natural to think that the features of the shower-to-shower distribution of the number of muons at the ground, $\Nmu$, contain information about the shower development. The distribution of muon number recorded at the ground is shown in Fig.~\ref{fig:scheme} (left) for showers initiated by proton with energy of $E = 10^{19}\,$eV and zenith angle of $\theta = 67^\circ$. The figure is shown for all the most commonly used hadronic interaction models: \qgs~\cite{qgsii}, \epos~\cite{epos-lhc} and \sib~\cite{sibyll}. It is interesting to note that this distribution resembles the distribution of the depth of the shower maximum, $\Xmax$. The most noticeable difference is that for the $\Nmu$ distribution, the exponential tail appears for low values while for the $\Xmax$ distribution it happens for high $\Xmax$ values. 
The number of muons at the ground depends on the shower inclination and observation level. While this can affect the $\Nmu$ distribution, it depends solely on the propagation of muons, which can be adequately taken into account~\cite{MuonTransport}. Hence, throughout the remaining of this manuscript, $\Nmu$ at the ground will be used as a proxy to the EAS muon content.
In~\cite{PLBalpha} it has been demonstrated, using EAS simulations that fluctuations of the $\Nmu$ distribution is substantially related with properties of the very first interaction, in particular, the ratio of energy that goes into \emph{hadronic} particles (see scheme in Fig.~\ref{fig:scheme} (right)). In fact, it has been observed that for all tested hadronic interaction models, there is a correlation of about 60\% between $\Ehad$ and $\Nmu$ as seen for instance in Fig.~\ref{fig:corr_alpha_nmu} (left). This correlation is stronger for low values of $\Ehad$ (and correspondingly for low $\Nmu$). For larger values, especially above $\Ehad > 0.8$, the correlation gets worse due to very high multiplicity events and diffractive interactions. Therefore, the correlation between $\Ehad$ and $\Nmu$ can be further improved if one takes into account the first interaction multiplicity. Specifically, this can be achieved by redefining $\Ehad$ as

\begin{equation}
\alpha_1 = \sum_{i = 1}^m \left( \frac{E_i^{\rm had}}{E_0} \right)^\beta
\end{equation}

where $E_i^{had}/E_0$ is the fraction of energy carried by each \emph{hadronic} particle with respect to the primary energy and
$m$ is the number of particles in the hadronic component of the interaction. The parameter $\beta$ is the power-law index that weights the growth of the average number of muons with the primary energy, $E_0$. From a simple Heitler-Matthews model~\cite{Heitler} it can be shown that $\beta$ is given by 

\begin{equation}
\beta = \frac{\log \left( m \right)}{\log \left( m_{tot}\right)}
\end{equation}

being $m_{tot}$ the total number of particles that emerge from the hadronic interaction. Hence, this parameter depends on the interaction multiplicity.
The improvement of the correlation of $\Nmu$ with this new quantity, $\alpha_1$ can be seen in Fig.~\ref{fig:corr_alpha_nmu} (right). The correlation factor between $\alpha_1$ and $\Nmu$ is greater than 0.7 for all post-LHC hadronic interaction models. It is shown in~\cite{PLBalpha} that $\alpha_1$ distribution of the first interaction is enough to describe the main features (the width\footnote{The distribution second moment.} and exponential tail) while the rest of the shower contributes only to the overall value of $\Nmu$. 
In fact, being the growth of the shower exponential and being the low-energy hadronic interaction much more numerous than the high-energy ones, it becomes evident that any deviation on the expected behavior at low energies would have a more significant impact on the average of the $\Nmu$ distribution.

While the above findings have been reached using simulation, it can be inferred by taking some considerations about the shower development. The production of muons in EAS arises nearly exclusively from the hadronic component of the shower. This means that the number of produced muons is closely related to the \emph{size} of the hadronic shower, i.e., the amount of energy that flows into this sector. Hence, the relevant quantity is the fraction of energy transmitted to the hadronic sector in each interaction, and not the multiplicity as one could naively think. 

Let us now consider that in the first interaction, only a small amount of energy was passed into the hadronic component. In this case, the size of the hadronic shower would be smaller, regardless of what might happen in subsequent interactions. Hence, the fluctuation of the number of muons that should depend more on the first interactions than on the later stages of the shower development. In fact, as pointed out above the fluctuations of the hadronic energy are so large and the number of sub-showers (multiplicity) so significant that the fluctuations of $\Nmu$ can be accounted for by the first interaction alone.

\section{Accessing the first interaction through the low-$\Nmu$ distribution tail}
\label{sec:3}
The possibilities attained to the plot in fig~\ref{fig:corr_alpha_nmu} (right) are enormous. For instance, in the case of a pure proton composition, the low values of $\Nmu$ would give access to the fraction of energy flowing to the hadronic component of the shower on the first interaction, while high values would be more sensitive to the interaction multiplicity.

As mentioned before, a striking feature of the $\Nmu$ distribution is its exponential tail at low $\Nmu$. This tail is more prominent for proton induced showers, and as the primary composition becomes heavier, the exponential behavior starts to disappear, and the $\Nmu$ distribution gradually become Gaussian (see Fig.~\ref{fig:calib} (right)). This behavior can be explained by a simple superposition model~\cite{PLBalpha}.

The above feature could be explored to measure the slope of the $\Nmu$ tail for proton induced showers, even in a mixed primary composition scenario, by employing a strategy similar to the one used to extract the proton-air cross-section analysing the $\Xmax$ distribution tail~\cite{PRLxsec}.

\subsection{Measurement of  $\Ehad$ distribution tail of the first interaction}
\label{sec:3.1}

In this section, we aim to demonstrate that the measurement of the slope of the tail of the $\Nmu$ distribution, $\Lambda_\mu$, within current experimental uncertainties, allows to assess a property of the multiparticle production of the first interaction. From Fig.~\ref{fig:corr_alpha_nmu} it can be seen that in the $\Nmu$ tail region, the variable $\Ehad$ dominates the behavior of $\Nmu$. Hence, given the strong correlation between these two quantities, the ansatz is that the measurement of $\Lambda_\mu$ is directly connected to the slope of $\Ehad$, which we will simply address as $\Lhad$.

\begin{figure}[t]
\centering
\includegraphics[width=0.45\textwidth]{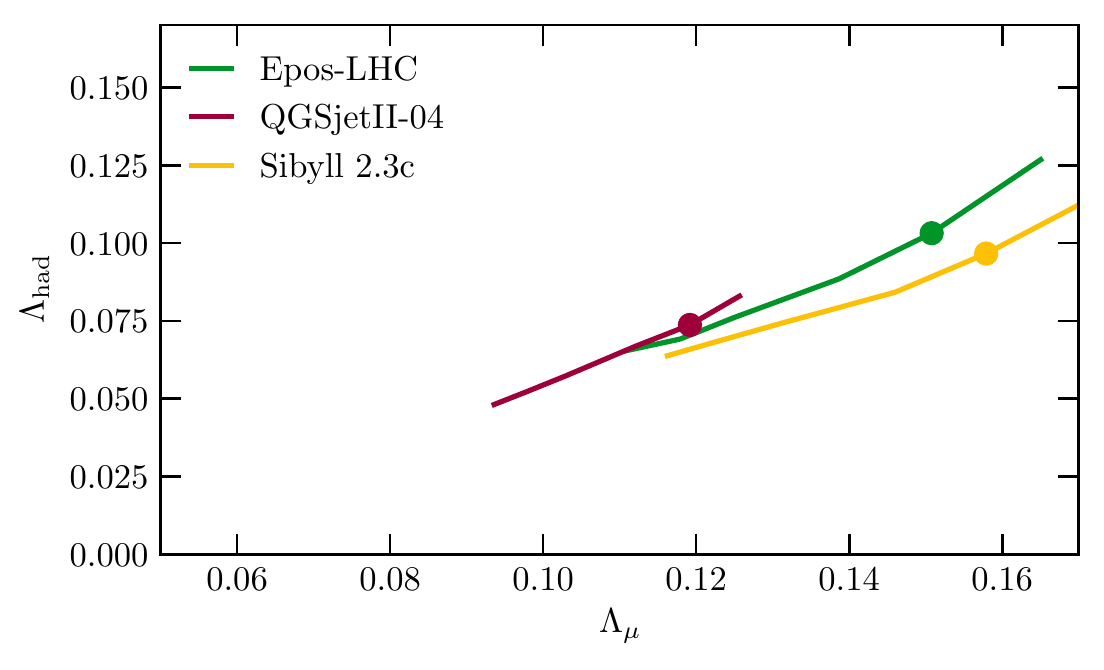}
\hfil
\includegraphics[width=0.45\textwidth]{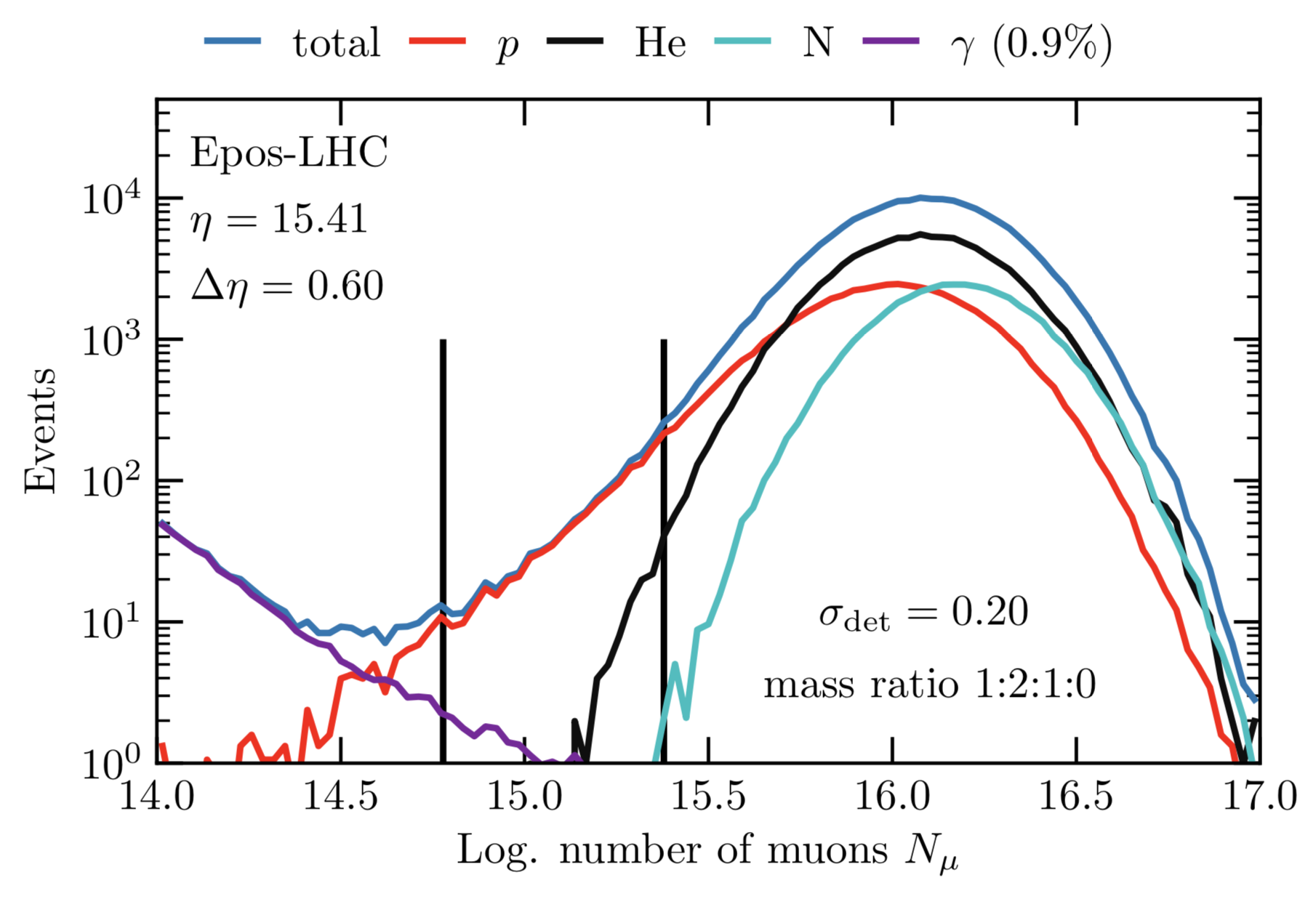}
\caption{(left) Conversion between $\Lambda_\mu$ and $\Lambda_\alpha$. The filled circles indicate the predictions by the different interaction models. The lines show how $\Lambda_\mu$ changes for each model if  $\Lhad$ is changed. (right) Number of showers with a given number of muons at ground for the nuclear primaries: p, He and N. The vertical bars indicate the fitting range.The total number of events is $\approx 200\,$k.}
\label{fig:calib}
\end{figure}

The measurement of $\Lhad$ has two major requirements: it should be possible to select a $\Nmu$ region where $\Lambda_\mu$ can be fitted, independently of the presence of heavier primary elements in the full distribution; there must exist a universal \emph{calibration} curve between $\Lambda_\mu$ and $\Lhad$.

The \emph{calibration} has been investigated using simulations of proton induced showers for different hadronic interaction models. The variable $\Lhad$ was changed artificially by selecting simulated showers from a large ensemble, and the impact on $\Lambda_\mu$ was then assessed.

The result of such exercise can be seen in Fig.~\ref{fig:calib} (left). In this figure, the points are the result for proton simulation while the curves are the result arising from this test. The results obtained show that a relation between $\Lambda_\mu$ and $\Lhad$ can be derived independently of the hadronic interaction models.

Now that we can interpret $\Lambda_\mu$ as $\Lhad$, it is necessary to prove that $\Lambda_\mu$ of the proton can be measured. For this, we have used the following mixed mass composition scenario: 25\% proton, 50\% helium, 25\% nitrogen, and no iron (or 1:2:1:0). This is one of the most pessimistic scenarios given by the analysis of the $\Xmax$ distribution in the Pierre Auger Observatory in the energy range of $log(E/eV) \in [18.5; 19]$~\cite{XmaxComp}. As the precision on $\Nmu$ at the ground is highly dependent on the shower energy, we decided, conservatively, to smear it by 20\%. This step aims to reproduce the most significant source of experimental uncertainty associated with the measurement of $\Nmu$.
Within this scenario, it is possible to see in Fig.~\ref{fig:calib} (right) that by choosing the fit region adequately it is possible to access the proton $\Lambda_\mu$ and ultimately measure $\Lhad$.

\begin{figure}[t]
\centering
\includegraphics[width=0.45\textwidth]{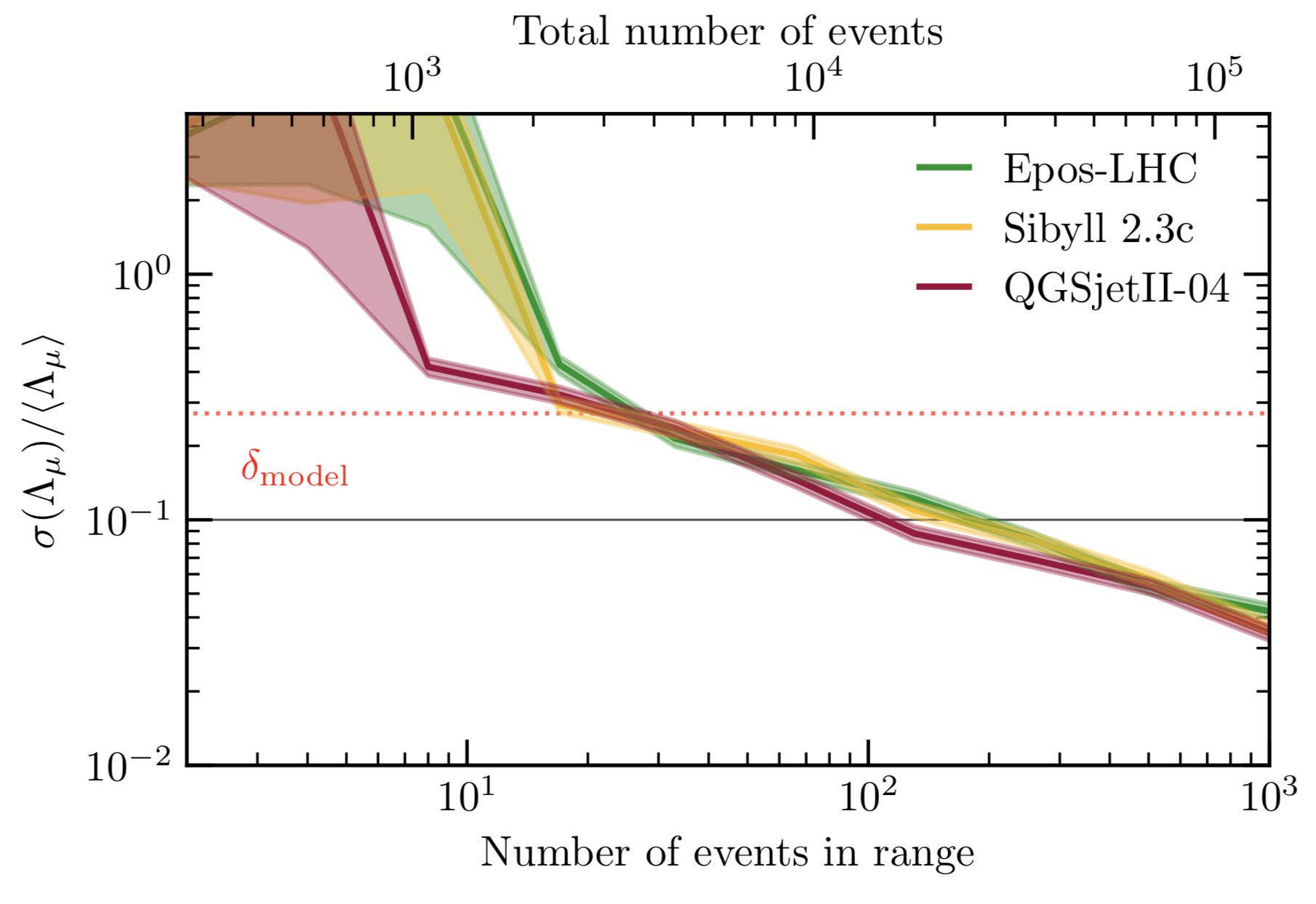}
\hfil
\includegraphics[width=0.45\textwidth]{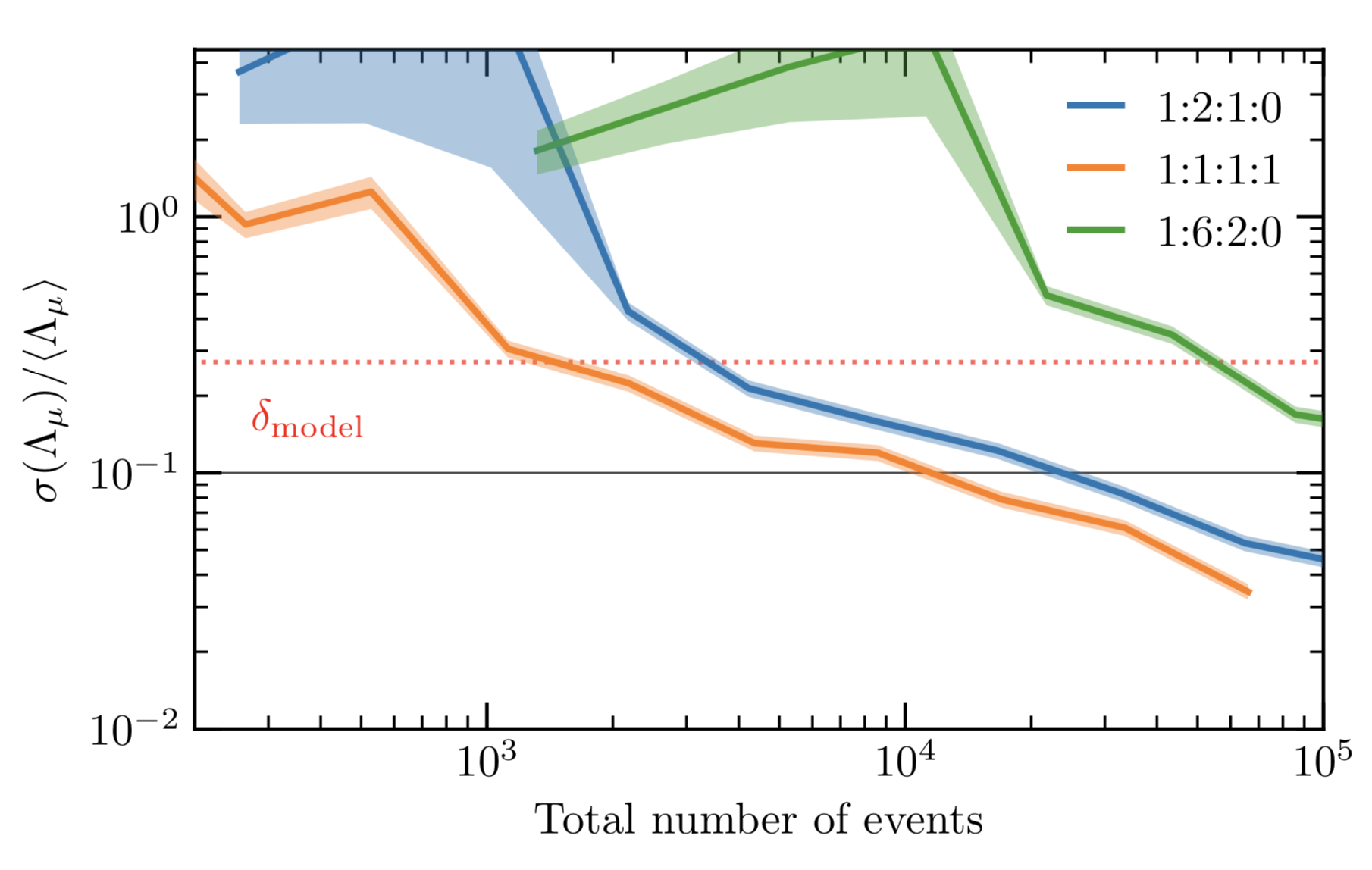}
\caption{(left) Relative fluctuation of the reconstructed $\Lambda_\mu$ as a function of the number of events in the experiments for the mass composition scenario 1:2:1:0. The x-bottom scale shows the number of events that fall in the fit region while the axis on top shows the total number of shower events collected.  The difference between the predicted slope for the different hadronic interaction models is 20\% (red dashed line). This level of precision is reached with 3000 events. The colored bands show the combined statistical error of the mean and standard deviation in the statistical sample. (right) Same plot as left generated with EPOS-LHC and for different mass composition scenario (see legend).}
\label{fig:tail_stat}
\end{figure}

The precision of the measurement of $\Lambda_\mu$ in the above-described scenario is shown in Fig.~\ref{fig:tail_stat} (left) for all tested hadronic interaction models. From this plot, it is possible to see that the precision of the measurement is only bounded to the number of collected shower events. The necessary precision to distinguish between hadronic interaction models using $\Lambda_\mu$ is marked in the plot as a dashed red line ($\delta_{model}$). For a $\Nmu$ distribution made of more than 3000 showers (or around 200 events in the fit range), the measurement precision becomes enough to discrimination between models.

Fig.~\ref{fig:tail_stat} (right) shows the same plot but for different mass composition scenarios. It is interesting to note that even in an extreme scenario with a huge amount of helium, such as 1:6:2:0, the measurement continues to depend solely on the number of recorded showers (even though one would have to increase the number of events with respect to the 1:2:1:0 scenario in one order of magnitude). 

The measurement of the slope of the tail of the $\Ehad$ distribution, $\Lhad$, is a quantity which characterizes the fluctuations on the energy flow of a hadronic interaction (in this case, the first one). For each interaction, there is a single $\Ehad$, which means that this it represents the overall behavior of the multiparticle production resulting from the hadronic interaction.

This does not make the quantity less appealing as it could be measured in accelerator experiments, for instance, in the forward experiments at the Large Hadron Collider, such as TOTEM~\cite{totem} and LHCf~\cite{lhcf}. Such overlapping measurement between accelerators and cosmic ray experiments is fundamental to confirm the validity of this analysis. For instance, LHC experiments are operating at a center-of-mass energy of 13 TeV. The equivalent energy for the cosmic ray interaction with the atmosphere nuclei is about $E \sim 10^{17}\,$eV. This is in an energy region that can be accessed by AMIGA scintillators~\cite{amiga}, at the Pierre Auger Observatory and therefore a cross-check could be performed. 
The above measurement would allow to strengthen the confidence in the measurement of $\Lhad$ at $E \sim 10^{19}\,$eV ($\sqrt{s} \sim 100\,$TeV), at an energy currently inaccessible to accelerator experiments.

\subsection{Access the neutral pion energy spectrum}
\label{sec:3.2}

While the measurement of $\Lhad$ of the first ultra-high energy interaction is already a significant breakthrough, it would be interesting to understand what are the multiparticle production properties, or interaction kinematical phase space, that is driving the slope of $\Ehad$.

The variable $E_{had}$ has direct relationship with the energy flow to the electromagnetic sector, $E_{em} = 1 - E_{had}$. Consequently, its fluctuations are the same, $\sigma(E_{em}) = \sigma(E_{had})$.

\begin{figure}[t]
\centering
\includegraphics[width=0.45\textwidth]{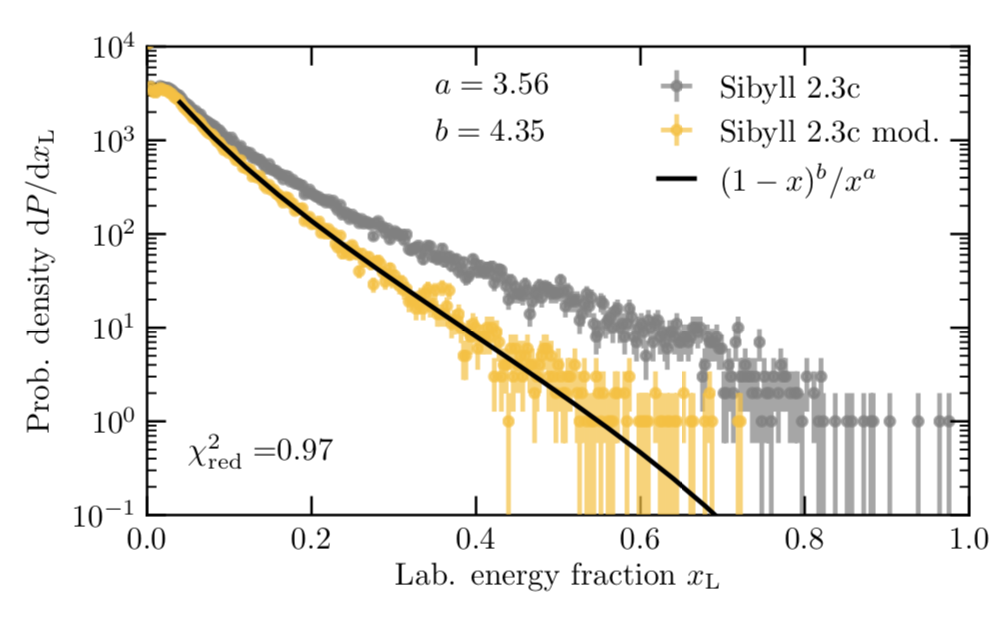}
\hfil
\includegraphics[width=0.45\textwidth]{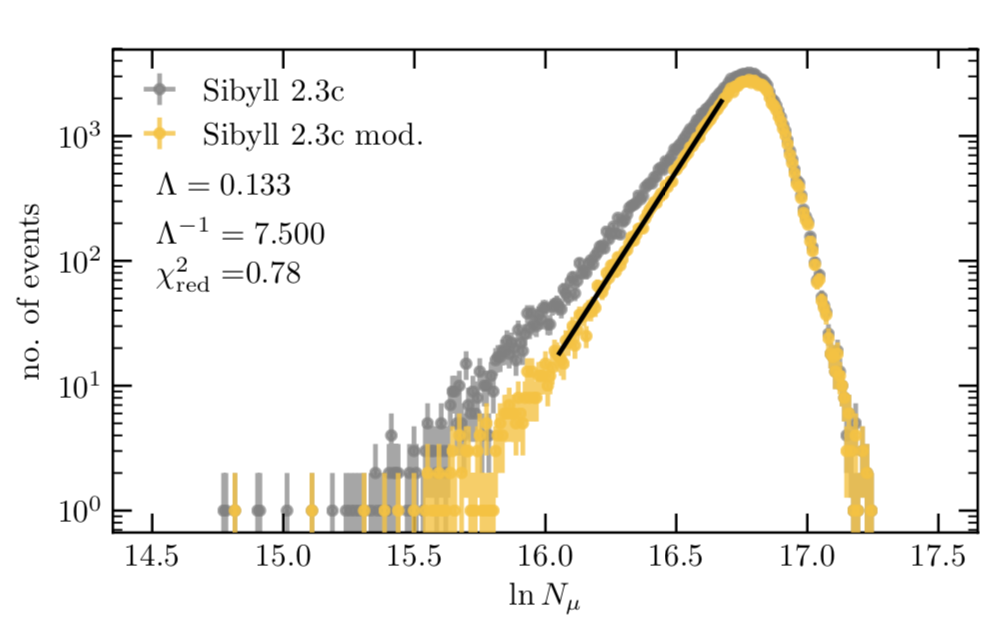}
\caption{(left) Energy spectrum of neutral pions as a function of the lab.\ energy fraction, $x_L$. The nominal cross-section in Sibyll 2.3c is shown in gray. The yellow curve represents a modified cross-section where production of neutral pions at large $x_L$ is suppressed. (right)  Distribution of the number of muons at ground in case of the two production spectra shown in the figure at the left.}
\label{fig:spectrum}
\end{figure}

The electromagnetic component is mainly fed by neutral pions\footnote{There is only an additional small contribution arising from excited resonances.}, whereas the hadronic component contains equal contributions from other mesons (such as charged pions and kaons ) and baryons (for instance protons). Therefore, it is natural to think that $\Ehad$ is connected with the energy spectrum of neutral pions. 

In fact, we have checked, using simulations, that a change on the high-energy tail of the  $\pi^0$ energy spectrum (inclusive cross-section as a function of the lab.\ energy), corresponds to a modification to the tail of the $\Nmu$ distribution, $\Lambda_\mu$, as shown in Fig.~\ref{fig:spectrum}. The result displayed in this plot strongly suggests that the measurement of $\Lambda_\mu$ gives access to a critical multiparticle production property: the behavior of fast partons in a hadronic interaction (including valence quarks).

From this exercise, one concludes that the tail of the $\Nmu$ distribution is highly dependent on the amount of energy that is given to the highest energy pions\footnote{The highest-energy particle is commonly referred as the \emph{leading} particle.}.

It has been checked that it is possible to derive a monotonous function that relates the slope of the tail of $\Nmu$ distribution with the high-energy tail of the neutral pions energy spectrum. However, contrary to $\Lhad$, this function has some dependence on the details of the hadronic interaction models. These are still preliminary studies, and the calibration might be improved through a more suitable choice of the energy range. 
However, it should be noted that even in the presence of a systematic uncertainty caused by hadronic interaction models, this measurement would allow to exclude new phenomena at the highest energies, namely violations to longitudinal scaling at high-rapidity, which is shared among all models or any other exotic new physics which would be revealed only at the $100\,$TeV scale.

\section{Conclusions}
\label{sec:conclusions}

In this work, it has been shown that the measurement of the features of the muon number distribution at the ground can be used to access properties of the first interaction of ultra-high energy cosmic rays.
In particular, it is shown that the exponential slope of the low-$\Nmu$ distribution can be used to measure the fraction of energy that flows into the hadronic component. Moreover, the slope of the $\Nmu$ distribution has also been found to be sensitive to the tail of the inclusive neutral pion production energy spectrum.

It was likewise proved, in this manuscript, that the measurement of the $\Lambda_\mu$ slope is possible within the current experimental uncertainties, being its major limitation the number of events. The analysis can be done even with an unknown mixed primary composition scenario, provided that there is a significant fraction of proton UHECRs.

The measurement of hadronic interactions at the highest energies might be relevant for the planning of future experiments such as the Future Circular Collider. Additionally, it can also provide an important handle to understand the shower development, solve the so-called muon problem, and ultimately get the mass composition of UHECRs unequivocally.

\bibliographystyle{jhep}
\bibliography{Biblio}

%

\end{document}